# Casimir force in the Gödel space-time and its possible induced cosmological inhomogeneity

Sh. Khodabakhshi[1], A. Shojai[1,2,a]

[1] Department of Physics, University of Tehran, Tehran, Iran
[2] Foundations of Physics Group, School of Physics, Institute for Research in Fundamental Sciences (IPM), Tehran, Iran

**Abstract** The Casimir force between two parallel plates in the Gödel universe is computed for a scalar field at finite temperature. It is observed that when the plates' separation is comparable with the scale given by the rotation of the space-time, the force becomes repulsive and then approaches zero. Since it has been shown previously that the universe may experience a Gödel phase for a small period of time, the induced inhomogeneities from the Casimir force are also studied.

## 1 Introduction

It was about half of the twentieth century that Gödel proposed an exact solution of Einstein field equations which is not in FLRW form and is compatible with an incoherent matter distribution [1]. A short description of the properties of the Gödel metric can be found in [2–4]. Interpreting the dust particles as galaxies, the Gödel solution could be a cosmological model of a rotating universe but this model exhibits no Hubble expansion [5]. Thus it cannot be a realistic model of our universe. Despite the unusual properties of the Gödel solution, in particular allowing the existence of closed time-like curves [6–9], it demonstrates a phenomenon that we cannot easily dismiss in general relativity. It is more rational to take it as an alternative metric which would in principle be allowed by general relativity and use the Gödel metric in studying the rotation of the universe.

One possible use of the Gödel solution is de Sitter–Gödel–de Sitter phase transition, a scenario in which the Gödel's rotatory property induces rotation in the universe during a phase transition from de Sitter to Gödel space-time and then back to de Sitter space-time [10]. In this scenario, the effective quantum potential of a scalar field, which may be the inflaton field, acts like a negative cosmological constant. It is shown that as the universe expands, it reaches a critical temperature in which the potential shows a negative valued minimum. Therefore it is possible to have a sequence of phase transitions from de Sitter to Gödel and back to de Sitter space-time. As is shown in [10], using the motion of a test particle and for a congruence of particles, it is seen that although the local induced rotation is of order of $\sqrt{\Lambda}c$, the global rotation is below the observational limit. This gives this scenario the ability to describe local rotation of galaxies without conflicting with the observed limit on the global rotation.

Assuming that the universe could make a rapid phase transition to Gödel space-time, it is fruitful to study more the local effects of this phenomenon. In the simulation of the induced rotation of the mentioned scenario, a uniform distribution of a large number (of order of $10^5$) of particles having a Gaussian distribution in the initial velocity has been considered and the result shows the local induced rotation is non-zero (see Fig. 6 of [10]). Dividing the space into adjacent cells, due to the fact that the quantum tunneling may happen with a probability, each cell may or may not experience the phase transition. In addition the symmetry axis of the Gödel space-time may be in any direction and this can generate some anisotropy. This produces local boundary layers, as explained later.

The universe is approximately homogeneous at large scales. The Cosmic Microwave Background (CMB) which is a snapshot of the oldest light in the universe, agrees well with the predictions of the $\Lambda$CDM model. Although it has a thermal black body spectrum of temperature $2.72548 \pm 0.00057$ K [11] it shows very small temperature fluctuations. Any model of the universe should explain these observations. In the study of the thermal history of the early universe [12–14], the CMB inhomogeneities are of wide interest. The standard approach to predict inhomogeneities is to consider the quantum fluctuation of a scalar field as the seed of perturbations and investigate their growth during the expansion of the universe [15,16].

[a] e-mail: ashojai@ut.ac.ir

It could be hypothesized that anisotropies generated by the de Sitter–Gödel–de Sitter phase transition could induce an inhomogeneity in the cosmic microwave background. Studying the possibility of such phenomena is the aim of this paper.

No one doubts the importance of the Casimir effect in modern physics. Since the Casimir prediction of a force between neutral conducting plates, which is an effect due to the vacuum fluctuations of quantum fields, a lot of research has been done both theoretically and experimentally on the Casimir effect [17–19]. Although the Casimir effect was originally predicted for electromagnetic fields it has also been calculated for scalar and other fields using different boundary conditions [20]. Besides studying the Casimir effect in flat space-time, investigating the Casimir effect in the curved space-time is increasingly making progress [21]. For cosmological models, to get closer results than the actual ones, one has to calculate the finite temperature Casimir energy [24,25].

An important property of the Gödel space-time is that it has an axis of symmetry. This leads to different values of Casimir force for different directions. Thus, in principle the Casimir force could induce an inhomogeneity in the cosmic microwave background, because it is direction dependent in Gödel space. We shall investigate such an effect in the following section.

In Sect. 2 we consider a scalar field living in the Gödel space-time subjected to the Dirichlet boundary conditions on two parallel plates and compute the Casimir force at finite temperature. Then in Sect. 3, using the obtained results and [10], the possibility of contributing to the inhomogeneities via the Casimir force is investigated.

## 2 Finite temperature Casimir force in the Gödel space-time

As is explained in the Introduction, recently we have shown [10] that the universe may experience a Gödel phase in a small period of time. Since this phase transition occurs locally, the Casimir force between the particles could produce some inhomogeneity in the early universe. Here we are interested in investigating the production of such inhomogeneities. In order to do this we first calculate the Casimir energy and force in Gödel background, using the method of effective action and zeta function regularization [22,23].

We consider the action functional of a massive scalar field coupled to a curved background, thus[1]:

---

[1] In general we can use the action $S[\phi] = \int d^4x \sqrt{-g} \left( \frac{1}{2} g^{\mu\nu} \partial_\mu \phi \partial_\nu \phi - \frac{m^2}{2} \phi^2 - \frac{\xi}{2} R \phi^2 - \frac{\lambda}{4!} \phi^4 \right)$. In the mean field approximation, the only difference is that in what follows, one should replace $m^2$ by $M^2 = m^2 + \left( \xi R + \frac{1}{2} \lambda \phi^2 \right)_{\text{mean field}}$.

$$S[\phi] = \int d^4x \sqrt{-g} \left( \frac{1}{2} g^{\mu\nu} \partial_\mu \phi \partial_\nu \phi - \frac{m^2}{2} \phi^2 \right). \tag{1}$$

In order to consider the quantum effects such as the Casimir effect, we can use the effective action given by

$$\Gamma_{\text{eff}} = -\ln \left( \int D\phi \, e^{-S[\phi]} \right) = \frac{1}{2} \text{Tr} \ln \left( \frac{\Box + m^2}{\mu^2} \right) \tag{2}$$

where

$$\Box = g^{\mu\nu} \nabla_\mu \nabla_\nu \tag{3}$$

and $\mu$ is an arbitrary parameter with dimension of a mass appearing in the regularization procedure. One of the best methods for the evaluation of the effective action is the zeta function method. Since the trace can be evaluated as the sum of the eigenvalues of the $(\Box + m^2)$ operator, it is useful to define the zeta function as

$$\zeta_{\left(\frac{\Box+m^2}{\mu^2}\right)}(s) = \sum_n \eta_n^{-s} \tag{4}$$

where $\eta_n$'s are the eigenvalues. Therefore the effective action is given by

$$\Gamma_{\text{eff}} = -\frac{1}{2} \zeta'_{\left(\frac{\Box+m^2}{\mu^2}\right)}(0). \tag{5}$$

To obtain the finite temperature effective action, we have to use the Euclidean action. The Casimir energy at finite temperature then can be obtained from the effective action [17]:

$$E_{\text{Casimir}} = -\frac{1}{2\beta} \zeta'_{\left(\frac{\Box+m^2}{\mu^2}\right)}(\beta, 0). \tag{6}$$

Choosing the background space-time to be the Gödel space-time, whose line element is given by

$$ds^2 = (dt + e^{\alpha x} dy)^2 - dx^2 - \frac{1}{2} e^{2\alpha x} dy^2 - dz^2, \tag{7}$$

where $\alpha$ is related to the intrinsic angular four-velocity by

$$\Omega^\beta = (0, 0, 0, \sqrt{2}\alpha), \tag{8}$$

and following the calculations of [26], the eigenvalues of the operator $(\Box + m^2)$ can be derived:

$$\eta = k_z^2 + m^2 + \alpha^2 \left[ n + \frac{1}{2} \right]^2 + \frac{1}{4}\alpha^2 - \left[ \omega + \sqrt{2}\alpha\varepsilon \left( n + \frac{1}{2} \right) \right]^2 \tag{9}$$

where $\varepsilon = k_y/|k_y|$. Note that the wave function is of the form $\exp(ik_y y + ik_z z - i\omega t)\psi(x)$ and the quantum number corresponding to the frequency is $n = 0, 1, 2, \ldots$.

Computations at finite temperature can be performed via the Matsubara replacement, in which the time dependence would be replaced with a periodic function. For Gödel spacetime, this leads to [26]

$$\left[\omega + \sqrt{2}\varepsilon\alpha\left(n + \frac{1}{2}\right)\right] \to \frac{2\pi i l}{\beta}, \quad (10)$$

in which $l = 0, \pm 1, \pm 2, \ldots$.

To obtain the Casimir energy, we assume we have two plates located in the $z$-direction, separated by a distance $d$. This yields the quantization of $k_z$ as follows:

$$k_z = \frac{n_z \pi}{d} \quad (11)$$

where $n_z = 1, 2, \ldots$

It is not necessary to be worried about the existence of $k_y$ in the eigenfunction; the eigenvalues are independent of $k_y$ and it can be integrated out like what have been done in [26]. Therefore the finite temperature $\zeta$-function would be

$$\zeta_{(\Box + m^2)}(\beta, s) = \sum_{l=-\infty}^{\infty} \sum_{n=0}^{\infty} \sum_{n_z=1}^{\infty}$$
$$\times \left[\frac{\pi^2}{d^2}n_z^2 + \alpha^2\left(n + \frac{1}{2}\right)^2 + \frac{4\pi^2 l^2}{\beta^2} + m^2 + \frac{\alpha^2}{4}\right]^{-s}. \quad (12)$$

In order to evaluate the zeta function, we shall write it in the form of a generalized Epstein (or Epstein–Hurwitz) multiple series, defined as[2]:

$$E_k(s; a_1, \ldots, a_k; c_1, \ldots, c_k; c)$$
$$\equiv \sum_{n_1, \ldots, n_k \in \mathbb{Z}} [a_1(n_1 + c_1)^2 + \cdots + a_k(n_k + c_k)^2 + c]^{-s}. \quad (13)$$

In doing this we notice that for small $\alpha$ the summand can be replaced by $\mathcal{A}$ defined by

$$\mathcal{A}(n, n_z, l) \equiv \alpha^{-2s}\left[\frac{\pi^2}{\bar{d}^2}n_z^2 + n^2 + \frac{4\pi^2 l^2}{\bar{\beta}^2} + \bar{m}^2 + \frac{1}{2}\right]^{-s}$$
$$= \alpha^{-2s}\bar{\mathcal{A}}(n, n_z, l) \quad (14)$$

where the dimensionless quantities are

$$\bar{\beta} = \beta\alpha, \quad \bar{d} = \alpha d, \quad \bar{m} = \frac{m}{\alpha}, \quad (15)$$

---

[2] For all $c_i = 0$, one should exclude $n_i = 0$ from the summation.

and we can rewrite the zeta function

$$\zeta_{(\Box + m^2)}(\beta, s) = \alpha^{-2s} \sum_{l=-\infty}^{\infty} \sum_{n=0}^{\infty} \sum_{n_z=1}^{\infty} \bar{\mathcal{A}}(n, n_z, l) =$$
$$\alpha^{-2s}\left(\frac{1}{2}{\sum_{n_z \in \mathbb{Z}}}' \bar{\mathcal{A}}(0, n_z, 0) + \frac{1}{4}{\sum_{n,n_z \in \mathbb{Z}}}' \bar{\mathcal{A}}(n, n_z, 0)\right.$$
$$\left. + \frac{1}{2}{\sum_{n_z,l \in \mathbb{Z}}}' \bar{\mathcal{A}}(0, n_z, l) + \frac{1}{4}{\sum_{n,n_z,l \in \mathbb{Z}}}' \bar{\mathcal{A}}(n, n_z, l)\right). \quad (16)$$

A prime over the summation means that the zero is excluded. In terms of Epstein series one obtains

$$\zeta_{(\Box + m^2)}(\beta, s) = \alpha^{-2s}\left(\frac{1}{2}E_1(s; a_2; c_2; c) + \frac{1}{4}E_2(s; a_1, a_2; c_1, c_2; c)\right.$$
$$+ \frac{1}{2}E_2(s; a_2, a_3; c_2, c_3; c)$$
$$\left. + \frac{1}{4}E_3(s; a_1, a_2, a_3; c_1, c_2, c_3; c)\right) \quad (17)$$

where

$$a_1 = 1, \quad a_2 = \frac{\pi^2}{\bar{d}^2}, \quad a_3 = \frac{4\pi^2}{\bar{\beta}^2},$$
$$c_1 = 0, \quad c_2 = 0, \quad c_3 = 0;$$
$$n_1 = n, \quad n_2 = n_z, \quad n_3 = l, \quad c = \bar{m}^2 + \frac{1}{2}. \quad (18)$$

Before proceeding, one should note that as usual [17] one should investigate the zero temperature ($\beta \to \infty$) separately. The normalized Casimir energy ($\bar{E} = E/\alpha$) will then be

$$\bar{E}_{\text{Casimir}}(\bar{d}, \bar{\beta}) = \bar{E}_0(\bar{d}) + \bar{\Delta}_{\text{F.T.}}(\bar{d}, \bar{\beta}). \quad (19)$$

$\bar{\Delta}_{\text{F.T.}}(\bar{d}, \bar{\beta})$ is the finite temperature part and $\bar{E}_0(\bar{d})$ is the zero temperature contribution defined by

$$\bar{E}_0(\bar{d}) = \lim_{\bar{\beta} \to \infty} \frac{-\zeta'(\bar{\beta}, 0)}{2\bar{\beta}}. \quad (20)$$

Introducing $\xi^2 = \frac{4\pi^2}{\bar{\beta}^2}l^2$, the zero temperature limit of the sum over $l$ could be replaced by an integral in the following way:

$$\lim_{\bar{\beta} \to \infty} \frac{1}{\bar{\beta}} \sum_{l=-\infty}^{\infty} f(\xi) = \int_{-\infty}^{\infty} \frac{d\xi}{2\pi} f(\xi). \quad (21)$$

All the relations can be simplified using the Epstein recursion formula [27],

$$E_k(s; a_1, \ldots, a_k; c_1, \ldots, c_k; c)$$
$$= \sqrt{\frac{\pi}{a_k}} \frac{\Gamma(s-1/2)}{\Gamma(s)} E_{k-1}(s; a_1, \ldots, a_{k-1}; c_1, \ldots, c_{k-1}; c)$$
$$+ \frac{4\pi^s}{\Gamma(s)} a_k^{-s/2-1/4} \sum_{n_1,\ldots,n_{k-1}\in\mathbb{Z}} \left[\sum_{j=1}^{k-1} a_j(n_j+c_j)^2 + c\right]^{-s/2+1/4}$$
$$\times \sum_{n_k=1}^{\infty} n_k^{s-1/2} \cos(2\pi n_k c_k) K_{s-1/2}$$
$$\times \left(\frac{2\pi n_k}{\sqrt{a_k}} \sqrt{\sum_{j=1}^{k-1} a_j(n_j+c_j)^2 + c}\right), \tag{22}$$

and the relations

$$K_{-1/2}(z) = \sqrt{\frac{\pi}{2z}} \exp(-z) \tag{23}$$

and

$$\Gamma(s) = \frac{1}{s} - \gamma + O(s). \tag{24}$$

For the zero temperature contribution after some lengthy calculations we have

$$\bar{E}_0(\bar{d}) = \text{divergent terms} - \frac{1}{4\pi} \sum_{k=1}^{\infty} \frac{e^{-2\pi kc/a_2}}{k} \int_{-\infty}^{\infty} d\xi e^{-2\pi k\xi^2/a_2}$$
$$- \frac{1}{8\pi} \sum_{n\in\mathbb{Z}}' \sum_{k=1}^{\infty} \frac{1}{k} \int_{-\infty}^{\infty} d\xi e^{-2\pi k\sqrt{(a_1 n^2 + c + \xi^2)/a_2}}. \tag{25}$$

The divergent parts have to be renormalized by counter terms and the parameter $\mu$, leaving the finite part:

$$\bar{E}_0(\bar{d}) = -\frac{\sqrt{a_2}}{4\sqrt{2\pi}} \text{Li}_{3/2}(e^{-2\pi c/a_2})$$
$$- \frac{1}{8\pi} \sum_{n\in\mathbb{Z}}' \sum_{k=1}^{\infty} \frac{1}{k} \int_{-\infty}^{\infty} d\xi e^{-2\pi k\sqrt{(a_1 n^2 + c + \xi^2)/a_2}}$$
$$\simeq -\frac{\pi}{24\bar{d}} - \frac{1}{4\sqrt{2\bar{d}}} \exp\left\{-2\left(\frac{1}{2} + \bar{m}^2\right) \frac{\bar{d}^2}{\pi}\right\} \tag{26}$$

where $\text{Li}_n(z)$ is the polygamma function.

The finite temperature part can also be simplified using the Epstein recursion relation:

$$\bar{\Delta}_{F.T.}(\bar{d}, \bar{\beta}) = \frac{1}{\bar{\beta}} \ln \left[\left\{1 - \exp\left(-\frac{2\left(\frac{1}{2} + \bar{m}^2\right)\bar{d}^2}{\pi}\right)\right\}^{\frac{1}{2}}\right.$$
$$\times \prod_{n_1,n_2,n_3,n_4=1}^{\infty} \times \left\{1 - \exp\left(-\bar{\beta}\sqrt{\frac{1}{2} + \bar{m}^2 + n_1^2 + \frac{\pi^2 n_2^2}{\bar{d}^2}}\right)\right\}$$

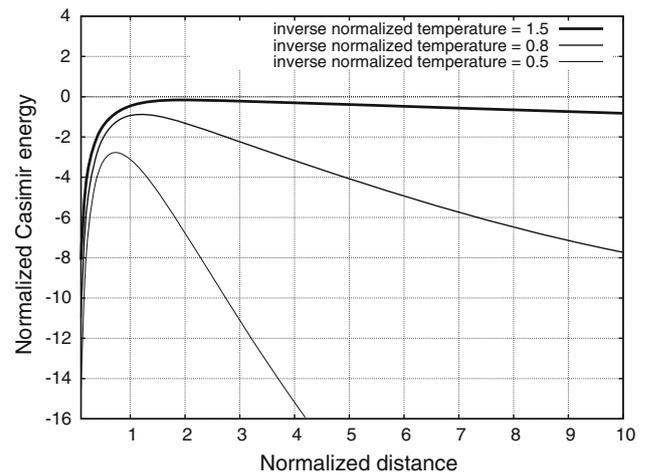

**Fig. 1** Normalized Casimir energy of a scalar field as a function of normalized separation for different normalized temperatures. $\bar{m}$ is set equal to 1

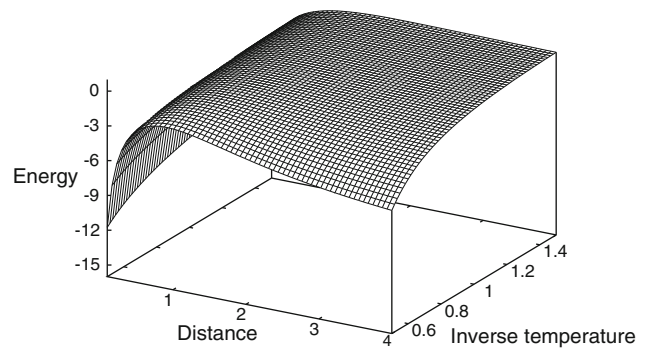

**Fig. 2** Normalized Casimir energy of a scalar field as a function of normalized separation and temperature. $\bar{m}$ is set equal to 1

$$\times \left\{1 - \exp\left(-\bar{\beta}\sqrt{\frac{1}{2} + \bar{m}^2 + \frac{\pi^2 n_3^2}{\bar{d}^2}}\right)\right\}$$
$$\times \left\{1 - \exp\left(-2\bar{d}\sqrt{\frac{1}{2} + \bar{m}^2 + n_4^2}\right)\right\}^{\frac{1}{2}}\right]. \tag{27}$$

The Casimir force can be derived from the Casimir energy via

$$F_{\text{Casimir}} = -\frac{\partial}{\partial d} E_{\text{Casimir}} = -\alpha^2 \frac{\partial \bar{E}}{\partial \bar{d}} = \alpha^2 \bar{F}_{\text{Casimir}}. \tag{28}$$

The Casimir energy and force are plotted in Figs. 1, 2, 3 and 4 as a function of normalized temperature and separation.

An interesting feature of the Casimir force in the Gödel space-time is that, in contrast to the flat space-time, the force somewhere near $\bar{d} = 1$ goes repulsive and for larger normalized distance approaches zero. This effect can be understood using a semiclassical argument. The Casimir force is in fact a result of the polarization of plates by trapping the virtual pairs created in vacuum (see Fig. 5a). The effect of Gödel space-time on the motion of particles is to introduce a helical motion with radius of order of $\alpha$. Considering the fact that the virtual particles live for a time given by the uncertainty relation, one observes that for a large value of the rotation factor (i.e. $\bar{d} \sim 1$) the virtual particles spend much of their time in spirals and cannot arrive at the plates. This is some depolarization effect that neutralizes the plates near $\bar{d} \sim 1$ and a small decaying repulsive force for larger distances.

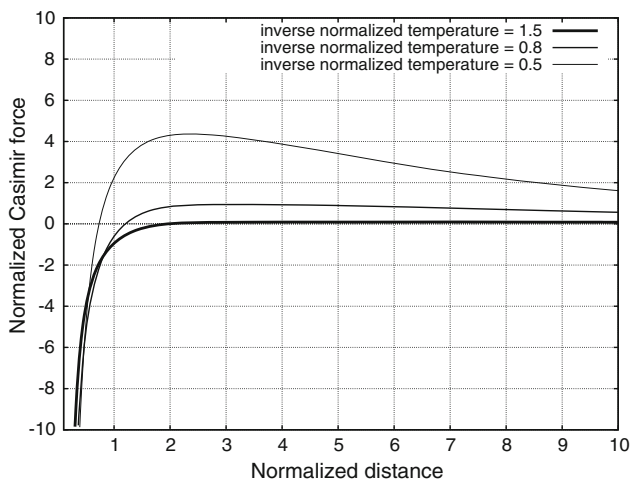

**Fig. 3** Normalized Casimir force of a scalar field as a function of normalized separation for different normalized temperatures. $\bar{m}$ is set equal to 1

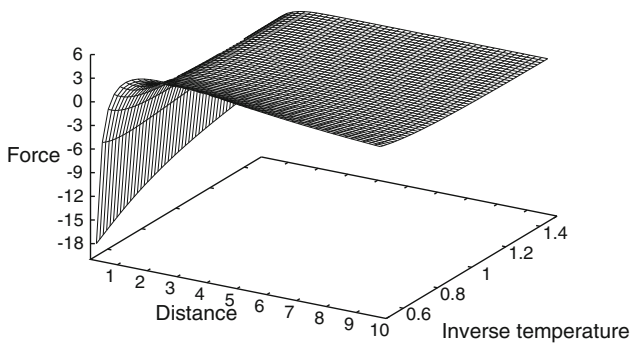

**Fig. 4** Normalized Casimir force of a scalar field as a function of normalized separation and temperature. $\bar{m}$ is set equal to 1

Finally, it should be noted that putting the boundary conditions in the $x$ or $y$ direction, there would be no noticeable change in the Casimir force with respect to the flat space-time case. This can be seen from the form of the eigenvalues. Therefore it is expected that some inhomogeneities would be induced into the universe. We shall investigate this in the next section.

## 3 Induced inhomogeneity from Casimir force in Gödel's background

As is stated previously, since the de Sitter–Gödel–de Sitter phase transition can happen locally [10], some inhomogeneity would be induced in the universe. This is because the Casimir force is sensitive to the direction of the rotation of the (local) Gödel space-time.

To investigate the Casimir force we have to clarify what the local boundary layers acting as boundary conditions are. It may be local layers of matter which could be produced in many ways. One way is to assume that these layers were produced by gravitational waves after the inflation era which can produce more dense regions.

But what is of our interest here is that the de Sitter–Gödel–de Sitter phase transition scenario creates a natural boundary condition. As explained in the Introduction, because of the fact that the quantum tunneling to Gödel space-time happens with a certain probability, it may or may not happen. As a result we can have local regions where such a transition is made or not in the neighborhood. The boundary between these neighboring regions would have different densities and act like the boundary layers for computation of the Casimir force. Since these neighboring regions are completely distinct, we can adopt Dirichlet boundary conditions.

As a result, local layers of matter would be redistributed in an inhomogeneous way. Since this phase transition may happen in the inflation era [10], the Casimir force could amend the inhomogeneities of the cosmic microwave background radiation.

In the de Sitter–Gödel–de Sitter phase transition scenario [10] the one-loop Euclidean effective potential of a scalar field plays the role of the cosmological constant. In the first phase (i.e. de Sitter phase), the effective potential acts like a positive cosmological constant and it is much larger than the dust density. After reaching the critical temperature, the phase transition happens and the effective potential plays the role of a negative cosmological constant which must be equal to $-\rho_{\text{dust}}/2$ to have the Gödel space-time. Finally the scalar field slowly rolls so that the universe goes to the de Sitter phase again. This process is shown in Fig. 6 [10].

As is shown in [10], the cooling of the universe that is needed to go to the Gödel phase is achieved via expansion and the critical temperature is given by the relation $\frac{d^2 V_{\text{eff}}}{d\phi^2}|_{\phi=0} = 0$. Then in the Gödel phase, slow rolling takes $\tilde{t}$ seconds, given by

$$\tilde{t} = \sqrt{\frac{\hbar G^2 \rho_{\text{dust}}}{\Lambda c^7}} = t_{\text{Planck}} \frac{\Lambda_{\text{Godel}}}{\Lambda_{\text{de Sitter}}} \quad (29)$$

where $t_{\text{Planck}}$ is the Planck time, $\Lambda_{\text{Godel}}$ and $\Lambda_{\text{de Sitter}}$ are the cosmological constant in the Gödel and de Sitter phase, respectively.

**Fig. 5** Virtual particles polarize the plates: **a** flat space-time, **b** Gödel space-time, with small $\bar{d}$, **c** Gödel space-time, with large $\bar{d}$

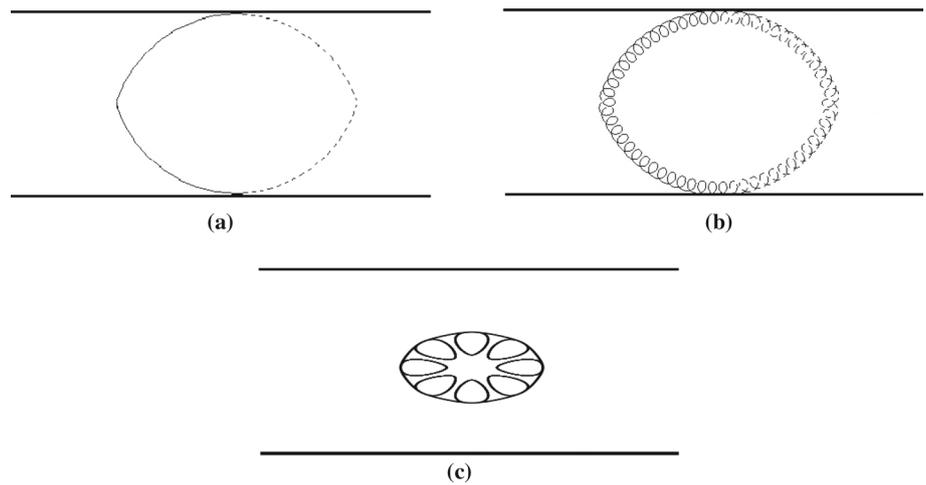

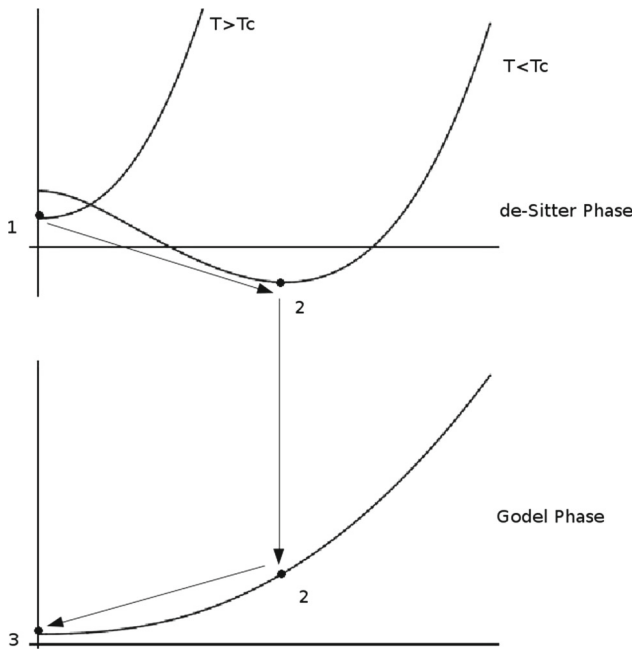

**Fig. 6** A typical scenario of the universe going through de Sitter, Gödel and de Sitter phases. See [10]

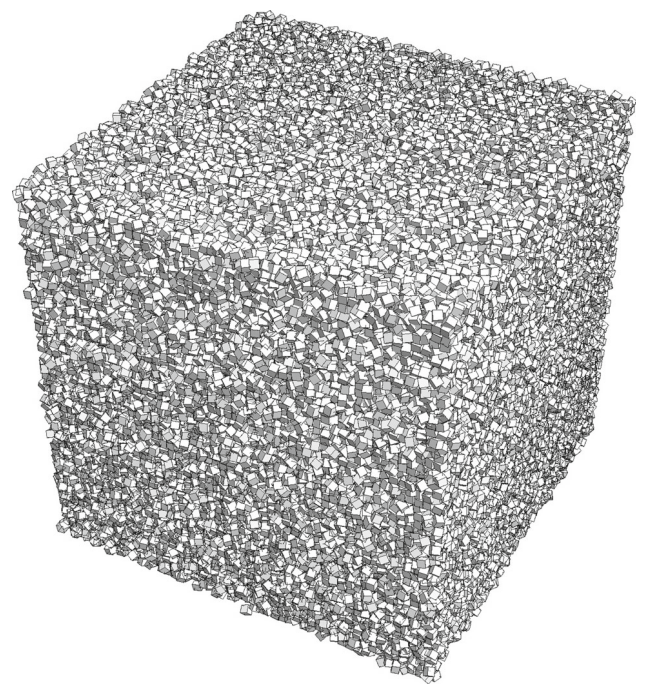

**Fig. 7** Simulation of the induced inhomogeneity

We have shown in [10] that being in the Gödel phase for $\tilde{t}$ seconds induces some amount of rotation in the motion of test particles which is of order $\sqrt{\Lambda}$. The induced rotation on a congruence of $1.5 \times 10^5$ particles is simulated in (Fig. 7) of [10]. The result shows that with such a mechanism a global rotation within the observational acceptable limit, while having local rotation, can be obtained.

Here we want to notice that during the Gödel phase the adjacent matter layers experience a Casimir force which, according to the results of the previous section, is sensible to the direction and this produces inhomogeneity.

Consider a cube cell with length, width and height $d$. It would shrink by an amount $\delta$ in the (random) z-direction because of the Casimir force during the Gödel phase. A computer simulation for a $50 \times 50 \times 50$ lattice of such cubes has been done. For each cell, it is assumed that the transition is such that the axis of the local Gödel phase is in a uniformly distributed random direction. Then using the Casimir force obtained the new sizes of the sides of each cell are calculated. The resulting matter distribution is shown in Fig. 7. As is clear from this figure, small perturbations to the global homogeneity would be produced.

The density of inhomogeneities can be estimated as follows. The maximum difference in the density that will occur after the Gödel phase is

$$\frac{\delta\rho}{\rho} = \frac{\frac{1}{d^3} - \frac{1}{d^2(d-\delta)}}{\frac{1}{d^3}}. \tag{30}$$

Taking $\delta$ small, we have

$$\frac{\delta\rho}{\rho} = \frac{\delta}{d}. \tag{31}$$

In order to compute the magnitude of $\delta$, we should solve the geodesic equation of a test particle under the Casimir force in the Gödel universe. We have

$$\frac{d^2z}{d\tau^2} + \Gamma^z_{\mu\nu}\frac{dx^\mu}{d\tau}\frac{dx^\nu}{d\tau} = \frac{F_{\text{Casimir}}(z)}{m}. \tag{32}$$

But because of the fact that $\tau = \tilde{t}$ is small (see [10]), the acceleration is nearly constant and the Newtonian limit will work well; $\delta$ is simply

$$\delta = \frac{1}{2m}F_{\text{Casimir}}(d)\tilde{t}^2. \tag{33}$$

Recovering all c, $\hbar$ and G's and noting that the force behaves like $1/\bar{\beta}\bar{d}^2$ for small $\bar{\beta}$ and $\bar{d}$, we get

$$\frac{\delta\rho}{\rho} \simeq \frac{1}{\beta d^4}\frac{\sqrt{\Lambda_{\text{Godel}}}}{\Lambda^3_{\text{de Sitter}}}\frac{1}{l_{\text{Planck}}m_{\text{Planck}}c^2}. \tag{34}$$

Using the fact that during the inflation, a number of e-foldings of order 65 is needed and that inflation happens between the time $t_i \sim 10^{-36}$s and $t_f \sim 10^{-32}$s, we have

$$\frac{c^2}{3}\Lambda_{\text{de Sitter}}(t_f - t_i) \sim 65; \tag{35}$$

and noting that the temperature is between $10^{26}$ K and $10^{30}$ K and the density is between $10^{-23}\frac{\text{kg}}{\text{m}^3}$ and $10^{-26}\frac{\text{kg}}{\text{m}^3}$, one observes that this gives density perturbations between $10^{-6}$ and $10^{-5}$. This is quite acceptable. Numerical simulation gives the same result.

It should be noted here that an important feature of the observed CMB temperature perturbations is that they are predominantly Gaussian. The quantum fluctuations are random and this suggests that one has a Gaussian distribution [28]. This is usually expressed in terms of the power spectrum index $n_S$. Gaussianity leads one to expect $n_S \simeq 1$. Observations show that $n_S = 0.960 \pm 0.014$ [29]. Our simple model clearly does not have the ability to explain the exact properties of the cosmological perturbations like Gaussianity.

In order to investigate the possibility of explanation of all the properties of the cosmic perturbations in this way, we have to simulate the de Sitter–Gödel transition considering the quantum tunneling probability. Then the resulting curvature perturbation and its growth during the expansion of the universe should be considered. In this way the predicted power spectrum of perturbations can be obtained. Here we have only shown that it is possible to get inhomogeneities from the de Sitter–Gödel–de Sitter transition scenario. Exact properties of the induced inhomogeneities will be investigated in a forthcoming work.

## 4 Conclusion

In the framework of our previously presented scenario[10], the universe may experience a (local) Gödel phase during the inflation era. Since the symmetry axis of this local Gödel space-time is randomly chosen via a spontaneous symmetry breaking mechanism, one expects that the Casimir force between layers of matter introduces some inhomogeneity. This is because of the fact that the Casimir force in Gödel space-time depends on the direction of the rotation axis of the local Gödel phase.

Here we calculated the Casimir force for a scalar field in the Gödel space-time using the techniques of finite temperature quantum field theory. It is observed that there is some induced inhomogeneity during the ending period of inflation and that its value is not far from the expected inhomogeneity.

**Open Access** This article is distributed under the terms of the Creative Commons Attribution 4.0 International License (http://creativecommons.org/licenses/by/4.0/), which permits unrestricted use, distribution, and reproduction in any medium, provided you give appropriate credit to the original author(s) and the source, provide a link to the Creative Commons license, and indicate if changes were made.
Funded by SCOAP[3].